\documentclass[useAMS,usenatbib,usegraphicx,a4]{article}
\usepackage{times}
\usepackage{url}
\usepackage{epsfig}
\usepackage{verbatim}
\usepackage[flushleft]{threeparttable}
\usepackage{amssymb}
\usepackage{amsmath}
\usepackage{pdfpages}
\usepackage{array}
\usepackage{comment}
\usepackage{hyperref} 

\usepackage{graphicx}
\usepackage{longtable}
\usepackage{ltablex}
\usepackage{threeparttable}
\usepackage{multirow}
\usepackage{rotating}
\usepackage[export]{adjustbox}
\usepackage{aas_macros}
\usepackage{authblk}

\newcolumntype{H}{>{\setbox0=\hbox\bgroup}c<{\egroup}@{}}

\def\deg{{^\circ}} 
\renewcommand{\apjl}{Astrophys. J. Lett.}   % Astrophysical Journal, Letters
\renewcommand{\apjs}{Astrophys. J. Suppl. Ser.}   % Astrophysical Journal, Supplement
\renewcommand{\aj}{Astron. J.}   % Astronomical Journal
\renewcommand{\apj}{Astrophys. J.}   % Astrophysical Journal
\renewcommand{\aap}{Astron. Astrophys.}   % Astronomy and Astrophysics
\renewcommand{\aapr}{Astron. Astrophys. Rev.}   % Astronomy and Astrophysics Reviews
\renewcommand{\mnras}{Mon. Not. R. Astron. Soc.}   % Monthly Notices of the RAS
\renewcommand{\nat}{Nature} % Nature
\renewcommand{\pasa}{Publ. Astron. Soc. Aust.}   % Publications of the Astron. Soc. of Australia
\renewcommand{\pasp}{Publ. Astron. Soc. Pac.}   % Publications of the Astron. Soc. of the Pacific
\usepackage[sort&compress,super,comma]{natbib}

\begin{document}

\newcommand{\mgii}{Mg\,\textsc{ii}}
\newcommand{\mgiil}{Mg\,\textsc{ii}$\lambda$2799}
\newcommand{\civl}{C\,\textsc{iv}$\lambda$1549}
\newcommand{\feii}{Fe\,\textsc{ii}}

\title{The accretion of a solar mass per day by a 17-billion solar mass black hole}

\author[1,2]{Christian Wolf (christian.wolf@anu.edu.au)}
\author[1]{Samuel Lai}
\author[1]{Christopher A. Onken}
\author[1]{Neelesh Amrutha}
\author[3]{Fuyan Bian}
\author[4]{Wei Jeat Hon}
\author[5]{Patrick Tisserand}
\author[4]{Rachel L. Webster}

\affil[1]{Research School of Astronomy and Astrophysics, Australian National University, Cotter Road Weston Creek, ACT 2611, Australia}
\affil[2]{Centre for Gravitational Astrophysics, Australian National University, Building 38 Science Road, Acton, ACT 2601, Australia}
\affil[3]{European Southern Observatory, Alonso de C\'{o}rdova 3107, Casilla 19001, Vitacura, Santiago 19, Chile}
\affil[4]{School of Physics, University of Melbourne, Parkville VIC 3010, Australia}
\affil[5]{Sorbonne Universit\'{e}s, UPMC Univ Paris 6 et CNRS, Institut d'Astrophysique de Paris, 98 bis bd Arago, F-75014 Paris, France}
% FOR PATRICK PLEASE USE instead::: Sorbonne Universit\’e, CNRS, UMR 7095, Institut d’Astrophysique de Paris, 98 bis bd Arago, 75014 Paris, France

%\usepackage{hyperref} 
%\hypersetup{colorlinks,citecolor=blue,linkcolor=blue,urlcolor=blue}

%%%%%%% IMPORTANT: We disable hyperlinks by default with this line, to avoid the error "\pdfendlink ended up in different nesting level" while writing.
%\hypersetup{draft}
%\begin{document}

\date{draft \today}
%\pagerange{\pageref{firstpage}--\pageref{lastpage}} \pubyear{2017}
\maketitle

\begin{abstract}
%(up to 150 words max)
%Quasars are supermassive black holes feasting on surrounding matter, which make them outshine their host galaxies. 
Around a million quasars have been catalogued in the Universe by probing deeper and using new methods for discovery. However, the hardest ones to find seem to be the rarest and brightest specimen. In this work, we study the properties of the most luminous of all quasars found so far. It has been overlooked until recently, which demonstrates that modern all-sky surveys have much to reveal. The black hole in this quasar accretes around one solar mass per day onto an existing mass of $\sim$17 billion solar masses. In this process its accretion disc alone releases a radiative energy of $2\times 10^{41}$~Watts. If the quasar is not strongly gravitationally lensed, then its broad line region (BLR) is expected to have the largest physical and angular diameter occurring in the Universe, and will allow the Very Large Telescope Interferometer to image its rotation and measure its black hole mass directly. This will be an important test for BLR size-luminosity relations, whose extrapolation has underpinned common black-hole mass estimates at high redshift.
%
%Quasars are supermassive black holes feasting on surrounding matter, which make them outshine their host galaxies. Around a million quasars have been catalogued in the Universe by probing deeper and using new methods for discovery. In this work, we study the properties of the most luminous of all quasars found so far. It has been overlooked until recently, which demonstrates that modern all-sky surveys have much to reveal. The black hole in this quasar accretes around one solar mass per day onto an existing mass of $\sim$17 billion solar masses. In this process its accretion disc alone releases a radiative energy that is equivalent to the output of 500 trillion Suns. If the quasar is not strongly gravitationally lensed, then its broad line region (BLR) is expected to have the largest physical and angular diameter occurring in the Universe, and will allow the Very Large Telescope Interferometer to image its rotation and measure its black hole mass directly. This will be an important test for BLR size-luminosity relations, whose extrapolation has underpinned common black-hole mass estimates at high redshift. 
\end{abstract}
%\begin{keywords}
%surveys -- galaxies: active -- QSOs: general -- methods: observational
%\end{keywords}
%\end{frontmatter}

\clearpage

%(up to 3,000 words max) 

In 1963, Maarten Schmidt identified the first quasar\cite{Schmidt63}, known as 3C~273. It appeared as a remarkably bright star of 12th magnitude, while its redshift suggested that it was among the most distant objects known in the Universe at the time. The two facts together implied an implausibly huge output of light, and ever since then, newly found quasars have impressed with their immense energy release from a small region of space. This could only be explained as gravitational energy being converted into heat and light within a highly viscous accretion disc around a supermassive black hole \cite{HF63,SS73} (SMBH). Quasars are signposts of fast growth in SMBHs on public display and allow the study of these growth processes.

Finding large samples of quasars then provides population and growth statistics to explain the origin of SMBHs in the Universe\cite{Soltan82,Boyle00,Ri06,TV17}. Generally, the most luminous quasars contain the fastest-growing SMBHs, although the relation between mass accretion rate and luminosity is affected by the mass and spin of the black hole as well as the structure and viewing angle of the accretion disc and disc winds\cite{Bardeen70, DL11, Dexter11, SN12, Ca18, Starkey23, Lai23}.

Today, around a million quasars are known\cite{Fl21}, although some specimens stand out from the crowd: in 2015, the ultra-luminous quasar J0100+2802 at redshift $z=6.3$ was identified\cite{Wu15} with a SMBH of 10~billion solar masses\cite{Eilers23}. In 2018, an even more luminous object\cite{Wo18} was found, J2157--3602 at $z=4.7$, with a SMBH of 24~billion solar masses\cite{Lai23}. Although their luminosity implies rapid growth, their existence is hard to explain: when black holes start from the remnant of a stellar collapse and grow episodically within the Eddington limit, they are not expected to reach the evident masses in the time from the Big Bang to the epoch of their observation, which has triggered a search for alternative scenarios\cite{Volon10,Volon15,Amar19,ZK21}. 

\begin{figure}
\begin{center}
\includegraphics[width=0.5\columnwidth]{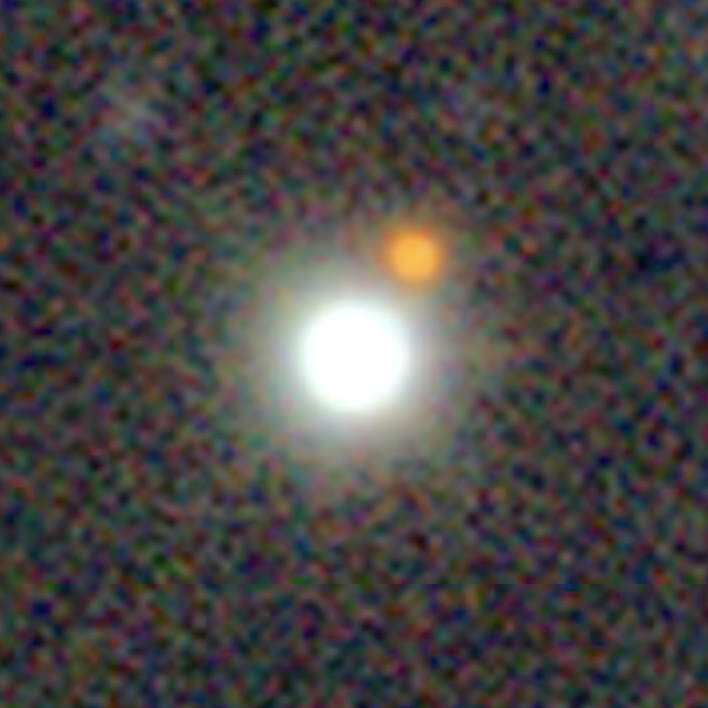}
\caption{The exceptional quasar J0529--4351: image from the Dark Energy Camera Legacy Survey DR10 with size $20\times 20$ arcsec$^2$; North is up and East is to the left; a neighbouring M star appears in red.}
\label{image}
\end{center}
\end{figure}

\begin{figure}
\begin{center}
\includegraphics[width=\columnwidth]{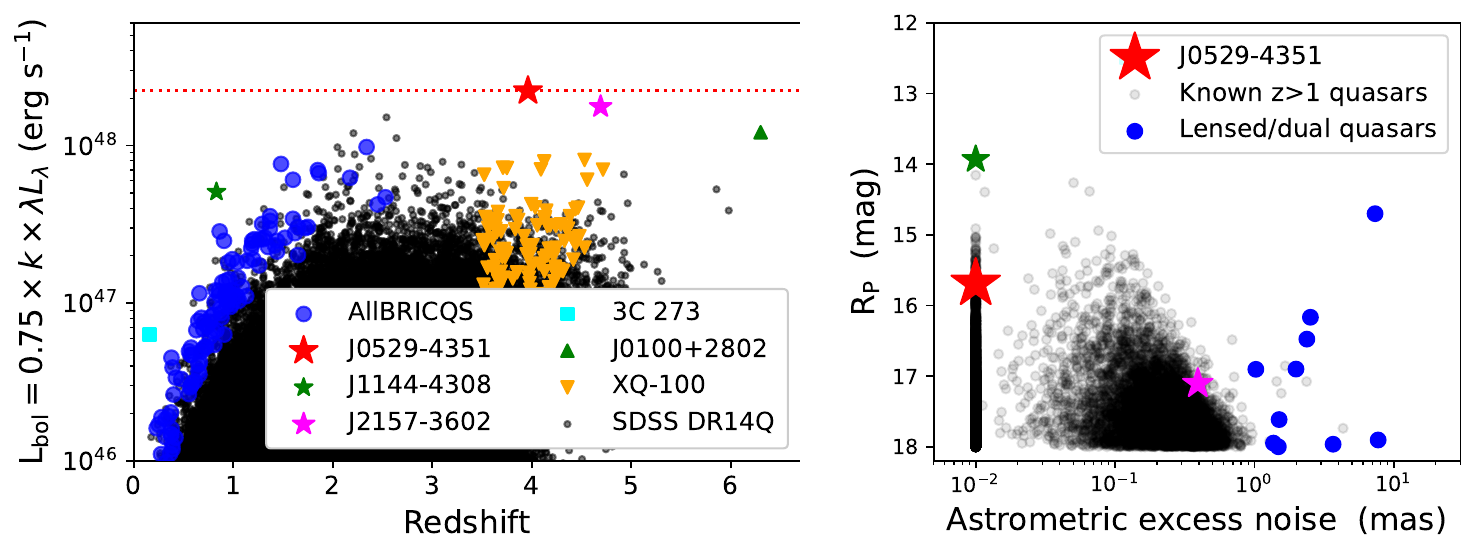}
\caption{J0529--4351 in relation to other quasars. Left: J0529--4351 vs the population from SDSS and from XQ-100, which has the highest-quality spectra\cite{Lai23b} for luminous quasars at $z \approx 4$. The luminosity axis shows monochromatic luminosity with standard bolometric corrections applied. Right: Astrometric excess noise (AEN) of J0529--4351 compared to known bright quasars at $z>1$; the known Gaia-unresolved lensed or dual quasars are all at AEN$>1$~mas. Zero AEN was replaced with a value of 0.01.}
\label{quasarpop}
\end{center}
\end{figure}

While exceptionally rare, the most extremely luminous quasars are interesting for several reasons beyond their intrinsic nature as discussed later. %further at the end. 
In this paper, we present the properties of the recently discovered\cite{On23} quasar SMSS J052915.80--435152.0, hereafter J0529--4351, which is a 16th magnitude object at redshift $z=3.962$ (see Fig.~\ref{image}), and reveal it to be the most luminous quasar currently known in the Universe (see Fig.~\ref{quasarpop}).

\section*{Results}

When quasars appear extremely bright, it may be suspected that their observed brightness is magnified by gravitational lensing from a massive galaxy on the line of sight. Strong lensing causes multiple separate images of a quasar in the sky\cite{Walsh79,Lemon23}. Two other quasars with a redshift and apparent brightness similar to J0529-4351 are known to be strongly magnified by lensing, the double-image APM~08279+5255 at $z=3.91$, with a separation of 0.5~arcsec\cite{Irwin98}, and the quadruply imaged B~1422+231 at $z=3.62$, with separations up to 1.5~arcsec\cite{Patnaik92}. Estimated magnification factors for these two objects range from 40 to 100, which implies that these quasars are not intrinsically extreme, but are members of the bulk population\cite{Irwin98,Egami00}.

J0529-4351, in contrast, shows no sign of strong lensing; data from the European Space Agency (ESA) {\it Gaia} satellite suggest it to be a point source, in terms of object morphology and astrometric excess noise, which has been used to find dual quasars or lensed quasars that appear unresolved to {\it Gaia}\cite{Chen22} (see Figure~\ref{quasarpop} and online methods for more detail). 
We also find no strong foreground absorber system, which our high signal-to-noise spectrum probes in \mgii\ at $z>1.15$; the strongest system, at $z=2.118$, has an equivalent width of EW(2796)=0.8\AA ; this suggests an impact parameter of 20~kpc (or 2.4 arcsec) from the line of sight\cite{Chen10} to the quasar, while a plausible image separation in a lensed scenario is $\sim 0.1$~arcsec.
We can also estimate the probability of lensing a source at $z\sim 4$ using common models for the galaxy distribution as isothermal masses\cite{Mason15,Yue22}, obtaining $p\approx 1.3\times 10^{-3}$. An image separation of 0.2 arcsec or less %from the Gaia observations, 
reduces it further to $p\approx 2\times 10^{-4}$. These estimates change by only a factor of a few when changing the input galaxy velocity dispersion function or the bright end of the quasar luminosity function. Even if a very steep intrinsic quasar luminosity function at the bright end enhanced the magnification bias, the probability that this source is strongly lensed will be less than 1\%.
We thus take the strength of the quasar emission at face value, although final confirmation from a high-resolution space-based or adaptive-optics image would still be desirable. 

\begin{figure}
\begin{center}
\includegraphics[width=\columnwidth]{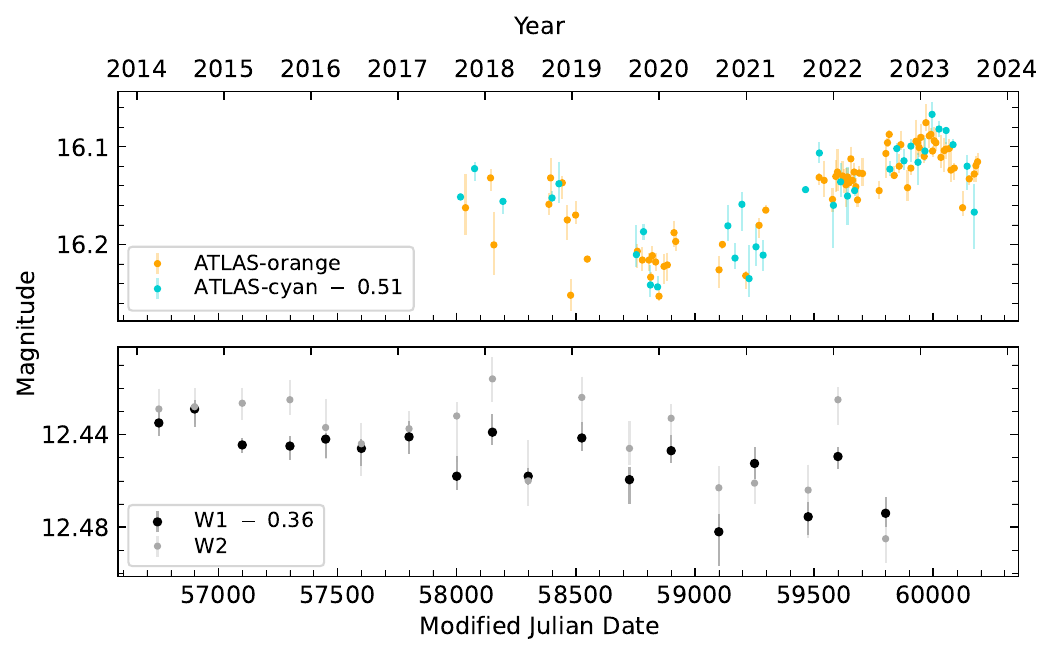}
\caption{Light curves of J0529--4351: data from NASA ATLAS are combined into weekly (orange passband) or moon-period (cyan passband) median averages of typically eight data points. Data from WISE are median averages of typically 25 visits completed within less than two weeks. Error bars show interquartile ranges within the averaging period. }
\label{LC}
\end{center}
\end{figure}

We also investigate the recent history in the brightness of J0529--4351 to see whether it may have been previously overlooked due to extreme variability. The 0.5-metre NASA (National Aeronautics and Space Administration) Asteroid Terrestrial-impact Last Alert System\cite{To18a} (ATLAS) telescope provides a light curve since mid-2017 (see Fig.~\ref{LC}). Brightness variations of $\sim 15$\% are found over the last six years, which are not unexpected for luminous quasars. Earlier records, from photographic plates observed in 1980 and 1998 (measured by the SuperCOSMOS Sky Surveys), found the $R$-band brightness to be consistent with recent observations by the SkyMapper Southern Survey\cite{Onken19} (SMSS). The somewhat longer light curve from the Wide-field Infrared Survey Explorer\cite{WISE} ({\it WISE}) shows also only modest variability.

The quasar is undetected in the Rapid ASKAP Continuum Survey\cite{McConnell20} (RACS), and hence has a flux of less than 1$\mu$Jy ($4\sigma$) in the broad 887~Mhz band. Using a common definition of radio loudness \cite{Kellermann89}, %as the flux ratio between 4~cm and 300~nm restframe wavelength, 
we find the object to be safely in the radio-quiet regime ($R<1$). Thus, we have no reason to suspect that its luminosity is affected by jet emission, let alone relativistically boosted.

\begin{figure}
\begin{center}
\includegraphics[width=0.9\columnwidth]{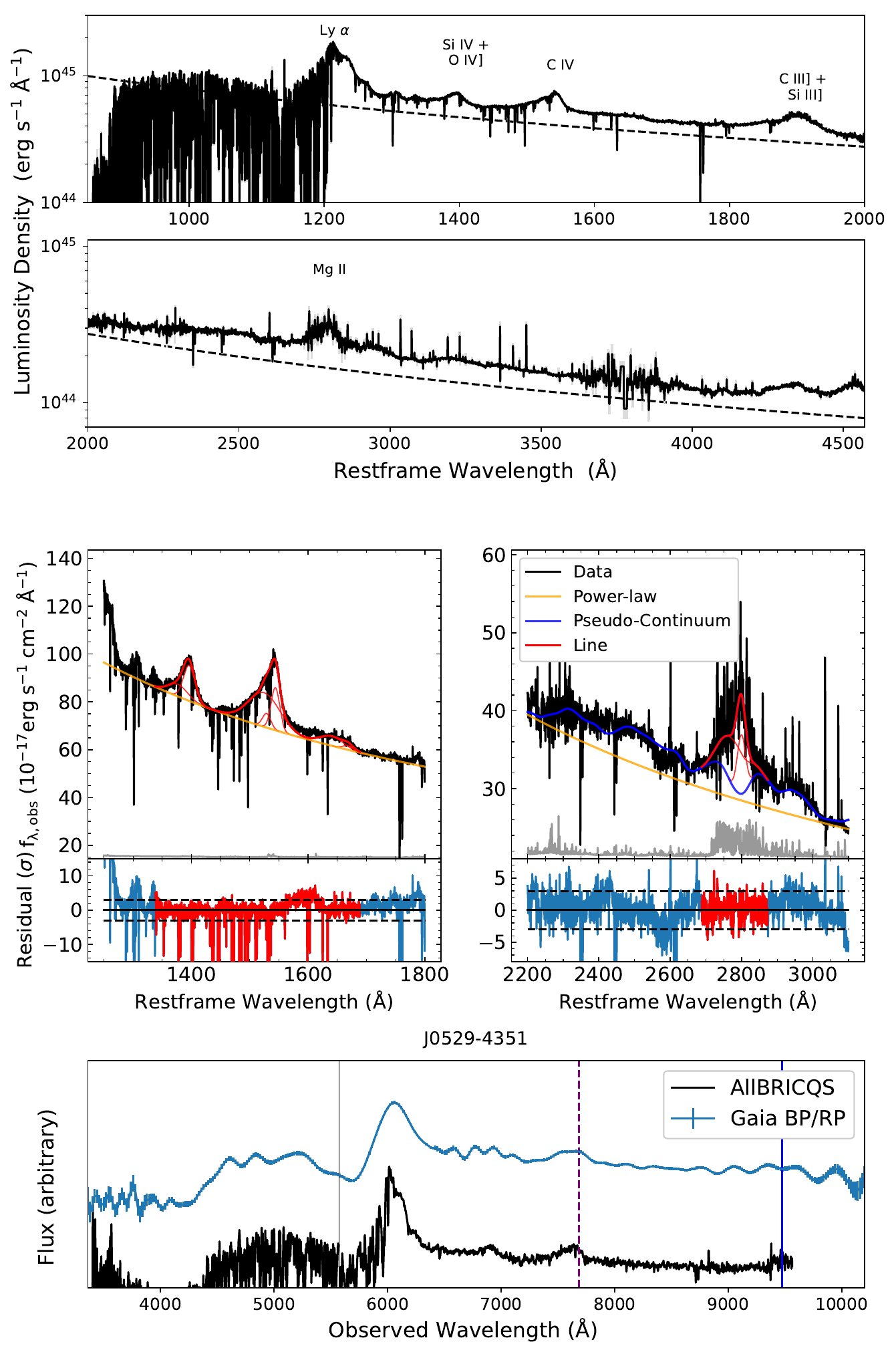}
\caption{Spectrum of the J0529-4351 from the ESO 8.2m Very Large Telescope. Top: observed-frame wavelength range of $0.42-2.27\mu$m; a power law with $\alpha_\nu=0.5$ is shown as a dashed line for comparison. Centre: decomposition, emission-line fits, and noise level (grey line, very low in the \civl\ part): derived black hole masses are $\log M/M_\odot=10.24$ for \civl\ (left) and $10.20$ for \mgiil\ (right). Bottom: earlier optical-only spectra from AllBRICQS and from Gaia DR3.
}
\label{spectrum}
\end{center}
\end{figure}

For further analysis, we use a spectrum of optical and near-infrared light (see Fig.~\ref{spectrum}) from %the 2.3m telescope of the Australian National University (ANU) and 
the 8.2m Very Large Telescope (VLT) at the European Southern Observatory (ESO). We split the spectrum with a publicly available spectrum-fitting code\cite{PyQSpecFit_v1} into an accretion disc continuum and emission-line contributions. From the disc continuum we quantify the disc luminosity of the quasar and get a first proxy for the accretion rate. We find monochromatic luminosities of $\log (L_{135}/\mathrm{erg~s}^{-1})=47.93$ and $\log (L_{300}/\mathrm{erg~s}^{-1})=47.76$ at restframe wavelengths of $\lambda=135$ and $300$~nm, respectively. An approximate estimate of the full radiative output from the accretion disc using standard bolometric corrections\cite{Ri06b} and a 0.75 anisotropy factor\cite{Runnoe2012} yields $\log (L_{\rm bol}/\mathrm{erg~s}^{-1})=48.37$. Using a standard radiative efficiency value\cite{YT02} of 0.1, this translates to an accretion rate of $\sim 413$~solar masses per year. This result makes J0529--4351 the most luminous quasar and by inference, the fastest-growing black hole in the Universe known to date (in terms of mass growth per unit time). 
%However, if the SMBH had maximum spin ($a=0.998$), the higher radiative efficiency of 0.32 would reduce the accretion rate to $\sim 132$~solar masses per year. 

\begin{figure}
\begin{center}
\includegraphics[width=\textwidth]{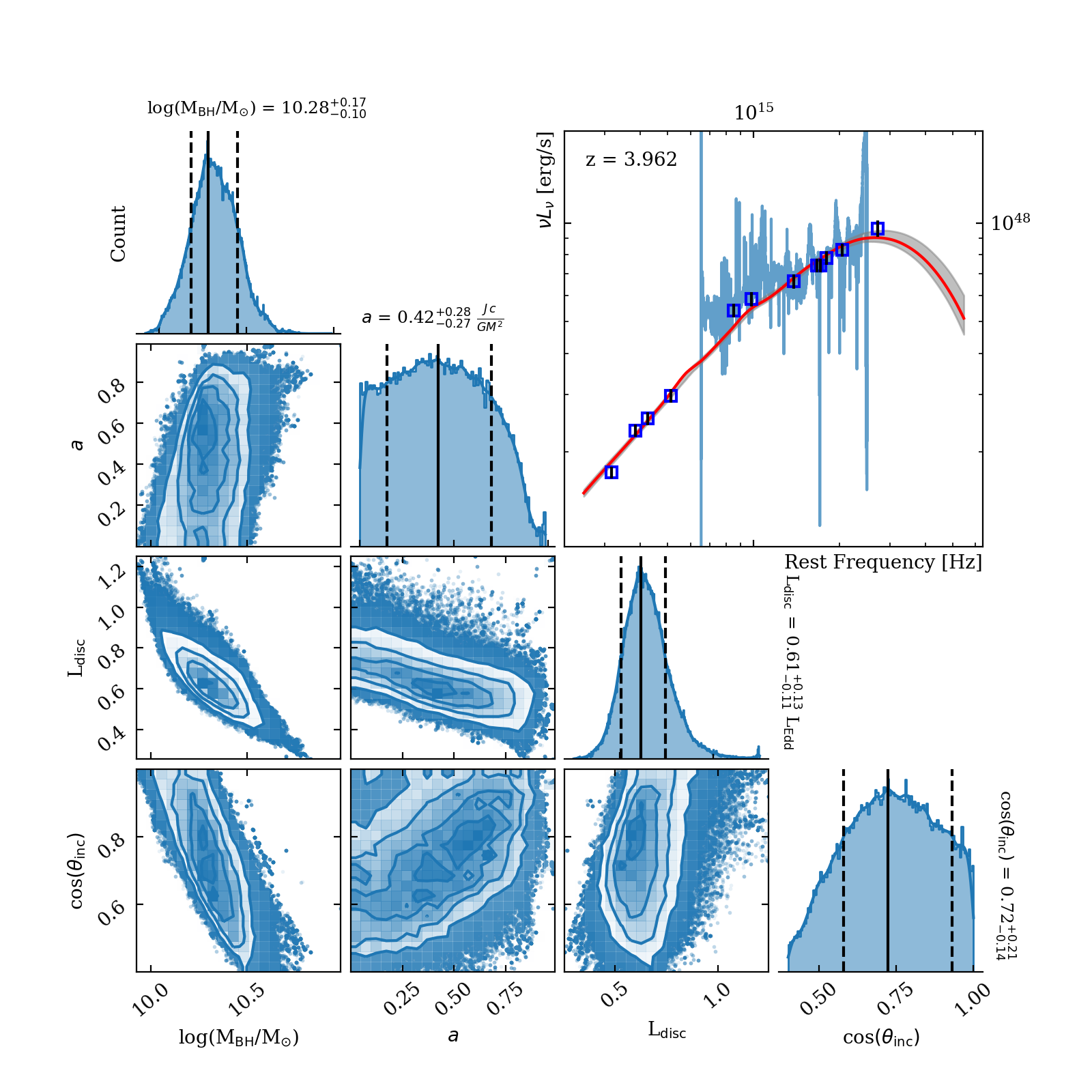}
\caption{Results of the continuum fitting for J0529--4351 using MCMC and slim disc models. Parameter estimates are given as peak of the probability distribution, and confidence intervals are central 68-percentiles. The highest likelihood model is a disc viewed at $\sim$45$^{\circ}$ angle around an intermediate-spin SMBH with $\sim$19~billion solar masses and an accretion rate of 370~$M_\odot$ per year; the constraints around viewing angle and spin are, however, weak. Spin is denoted by $a$ and L$_{\rm disc}$ is in units of Eddington luminosity.}
\label{ADfit}
\end{center}
\end{figure}

A more refined analysis of the bolometric luminosity and accretion rate requires modelling the possible spectra of the disc continuum over a grid of black-hole masses and spins and simulating their observation for different accretion rates and viewing angles\cite{Cald13,Cap15,Mejia18,Ca20}. Using slim-disc models as appropriate for SMBHs with high accretion rates\cite{Ab88,Sad11} and a publicly available code\cite{Lai23,BADFit_v1} (see online methods), we find a best-fitting solution to our spectrum being an Eddington-accreting disc around an SMBH with a moderate spin of $a \approx 0.4$ viewed at an intermediate angle of $i\approx 45\deg$ (see Fig.~\ref{ADfit}). The intrinsically emitted bolometric luminosity is $\log (L_{\rm bol}/\mathrm{erg~s}^{-1})=48.16$, which is 62\% of the value derived above; the lower value mostly results from generic bolometric corrections being overestimates for very massive black holes with their colder discs and suppressed UV emission. The best-fit slim disc model yields an accretion rate of 370~$M_\odot$ per year and a radiative efficiency of $\sim 0.09$.
%, and partly due to the nearly face-on viewing angle that enhances brightness relative to the average expected in a randomly orientated sample. 
%This new luminosity estimate corresponds to an accretion rate of $255$~solar masses per year. 
%In contrast to the case of J2157--3602\cite{Lai23} and many other quasars, the spectrum of J0529--4351 is not a very good fit to any of the thin or slim disc models we tried (see methods). 
Given the lack of higher-frequency data in the SED, the integrated model SED may still underpredict the bolometric flux, while the standard bolometric corrections must be overcorrecting for black holes of over a billion solar masses. The best estimate is expected to lie between the two, and we thus choose the average of $\log (L_{\rm bol}/\mathrm{erg~s}^{-1})=48.27$.  

%Most of the uncertainties in the model fit are due to degeneracies that cancel out with respect to the bolometric luminosity or accretion rate, which are mainly affected by the viewing angle. %This implies that luminosity and accretion rate may be up to 26\% lower in the extreme case of a pole-on view or instead over 35\% higher for $i>60\deg$. 
When modelling SEDs for a fixed mass accretion rate, different black-hole spins lead to different bolometric corrections and radiative efficiencies, while the monochromatic UV-optical luminosities are only modestly affected. The main uncertainty arises from the unknown viewing angle, which affects both the apparent monochromatic luminosity and the derived accretion rate.
Given the broad confidence intervals for spin and viewing angle in the model fit, we consider the full plausible angle range from pole-on to $i=60\deg$ as a 95\% error margin; this corresponds to a 95\% range in luminosity of $\pm 0.12$~dex and in accretion rate of 280 to 490 solar masses per year. 

%\textcolor{red}{Add a paragraph on the mean luminosity, final accretion rate estimates, and uncertainties based on unknown inclination? Mean is} $\log L_{\rm bol}=48.31 \pm 0.11$ \textcolor{red}{if I take the two k-corrections as independent measurements, which is what we decided on for the black hole masses.}  
%CW: I did not count the two passband-based ones as independent - we could define any number N of passbands without adding weight to this argument. 

\section*{Black-hole mass}

We estimate the mass of the black hole powering this quasar using two fundamentally different methods:

(1) Assuming that the continuum emission is affected at its blue end by the inner truncation of the accretion disc due to the innermost stable circular orbit around the black hole\cite{Bardeen70}, we can infer a combined estimate for mass and spin of the black hole. More massive black holes impose larger truncation radii %holes 
and move the peak of the continuum emission to cooler temperatures and longer wavelengths\cite{LD11}. With this method the mass of the black hole in 3C~273 was found to be between 200 and 500~million solar masses\cite{Malkan83}, 35 years before a value of $\sim$300~million solar masses was measured by interferometric observations\cite{VLTGRAV18} of the broad-emission line region in 3C~273. Since then, the continuum-fitting method has not only proven useful for estimating masses using thin disc models\cite{Cap15,Mejia18,Ca20} but has also been applied\cite{Lai23} with slim discs models\cite{Ab88} that are expected to be a more realistic description of the near-Eddington accretion discs of fast-growing SMBH. From the continuum shape of J0529--4351, we find a mass of $\log M/M_\odot=10.28^{+0.17}_{-0.10}$ (or $\sim 19$~billion solar masses).  

(2) Assuming that the broad emission-line region represents virialised gas moving at the velocity of Keplerian orbits around the SMBH, we can use the width of emission lines and the continuum luminosity to infer the mass of the black hole. This method is known as the virial single-epoch method\cite{MJ02,VP06} and has also been used to estimate the mass of the SMBH in J0100+2802\cite{Eilers23} at $z=6.3$. Its application at the high-luminosity end of the quasar population is mainly limited by a lack of independent calibrations for the SMBH mass and relies on extrapolations from lower-luminosity quasars calibrated with reverberation mapping\citep{Pet04}.

Our spectra offer two emission lines for this method, the triple-ionised carbon line \civl\ and the singly-ionised magnesium line \mgiil. The \civl\ line appears asymmetric and blueshifted relative to the \mgiil\ line ($\Delta v=-3120\pm80$~km/s) and a line full-width at half maximum (FWHM) of $7245\pm 175$~km/s. 
%We note that quasars with high \civl\ blueshift generally have continuum shapes\cite{Temple_2021} similar to J0529 that deviate from the simplest disc models and show a steeper blue end despite otherwise established high black hole masses. Thus, the spectrum of J0529 is not extraordinary and the continuum fitting method will remain unreliable for high \civl\ blueshift objects until their SEDs are better understood. 
The \mgiil\ line turns out to be hard to calibrate at this redshift because of atmospheric absorption, but we do measure a FWHM of $4,395\pm 435$~km/s. Using calibrations commonly applied to high-luminosity quasars, the line properties and luminosity translate into SMBH mass estimates from $\log M/M_\odot=10.03\pm 0.06$ to $10.45\pm 0.02$ for the CIV line\cite{Coatman_2017,VP06} and from $\log M/M_\odot=10.03\pm 0.09$ to $10.36\pm 0.09$ for the MgII line\cite{Vestergaard_2009,Shen_2011} (errors are standard deviations due to propagated observational uncertainties, and the scatter among the values is in line with large systematic calibration uncertainties). Table~\ref{tab:masstab} summarises all our estimates of masses and luminosities.

The line-based estimates are consistent with each other and with the continuum-based estimate. Given that the systematic uncertainties in these methods are larger than the statistical error propagation and may be as high as 0.4~dex, we combine the mean mass estimates of the two lines and that from the SED without weighting into a final result for the mass of $\log M/M_\odot=10.24\pm 0.02$ (simple mean and standard deviation), or $\sim 17$~billion solar masses. This also implies that the SMBH is accreting near the Eddington limit (Eddington ratio of $\sim 0.9$).

\section*{Discussion}

In terms of luminosity and likely growth rate, J0529-4351 is the most extreme quasar known. The accretion of J0529-4351 is near the Eddington limit, which is common among quasars of the highest luminosity\cite{Shen2008, Wu15, Lai23b}. The growth rates are mostly uncertain due to the unknown viewing angle. 
%A minor caveat is also that in a windy disc, as indicated by the asymmetric \civl\ line with a strongly blueshifted tail, the question whether the accretion rate suggested by the disc luminosity corresponds really to the black-hole accretion, and how much of the mass flow through the disc is lost to a wind before reaching the inner edge of the disc\cite{SN12}. This point might be especially relevant for tails seen in this object.
Assuming persistent accretion at its current Eddington ratio, the mass doubling time is $\sim 30$~Myr. However, with 19 billion solar masses, the black hole in J0529-4351 at $z=3.962$ is not the largest SMBH found in the most luminous quasars. It has over 50\% more mass than the black hole in J0100+2802 at $z=6.3$ but one third less mass than that in J2157-3602 at $z=4.7$. The age of the Universe at these three redshifts is 859~Myr, 1,244~Myr and 1,530~Myr, assuming a flat Universe with a concordance cosmology (a cosmological constant of $\Omega_\Lambda=0.3$). Given that J0529-4351 is observed at a later epoch in the Universe than J2157-4351 and J0100+2802, it is less of a challenge to models of early SMBH growth. 

%Choosing thin or slim disc models makes some difference for the estimate of its luminosity and accretion rate, although the thin disc model is less appropriate for discs with a high accretion rate. 

 %However, it is not necessarily typical for quasars with the most massive SMBHs, and this distinction is driven by the trivial propagation of luminosity into Eddington ratio. Fig.~\ref{quasarpop} contains several more massive black holes that are accreting much less. 
%We find no evidence for super-Eddington accretion, so the extreme luminosity is simply the result of a high black-hole mass combined with an accretion rate that is correspondingly large within naive Eddington constraints.

%The preference for a face-on viewing angle is not surprising given the object we study, because such viewing angles enhance observed brightness. It is interesting that the model fit prefers a low-spin black hole: 
It may be tempting to speculate on the spin of extreme SMBH: while it has been argued that growing black holes should spin up over time, this would also increase radiative feedback and slow down accretion, making it harder to grow the most massive SMBH within the age of the Universe; instead, one way for growing black holes from stellar seeds to the greatest masses we measure is ``chaotic accretion'' with randomly changing orientation that keeps the black-hole spins and radiative feedback low\cite{ZK21}.

A long-standing question has been what mechanism fuels the high accretion rate, which also must have persisted for some time already, though not necessarily in the immediate past. Mass and kinematics of gas in the host galaxy of J0529-4351 could be observed with the Atacama Large Millimetre Array (ALMA). ALMA has already revealed the largest spiral galaxy in the early Universe\cite{Tsukui21}, in the quasar BRI~1335-0417 at $z=4.4$, which is $\sim 10\times$ less luminous than J0529-4351. In contrast, ALMA observations have revealed a merger signature in the host galaxy of a quasar at $z=7.54$\cite{Ban19}. Further observations of extremely luminous quasars are progressing\cite{WISSH}. If extreme quasars were caused by unusual host galaxy gas flows, ALMA should see this; if nothing unusual was found in the host gas, this would sharpen the well known puzzle of how to sustain high accretion rates for long enough to form such extreme SMBHs.  %may either be a short episode in the life of the quasar related to nuclear events, .

\section*{Have we overlooked still more extreme quasars?}

%Interestingly, the most luminous quasars are not being discovered by pushing deeper into the Universe with more advanced facilities. After all, a 16th magnitude object is accessible to the CCD cameras of amateur astronomers with backyard telescopes (a spectrum of J2157-3602 has been observed by Rolf Wahl Olson with a 32-cm telescope). Instead, their nature presents us with a needle-in-the-haystack problem and they are revealed by using a more complete characterisation of their appearance. Such objects are often hiding in plain sight, such as the quasar J1144-4308 at redshift $z=0.83$, which was recognised as one of the brightest quasars in the night sky only in 2022, despite having been imaged on photographic plates since the 19th century\cite{On22}. 

Finding rare and exceptionally bright quasars such as J0529-4351 does not require large telescopes, but is instead a needle-in-the-haystack problem that needs precise data with discriminative power across large areas of sky. Such objects are often hiding in plain sight and are mistaken for stars. E.g., the quasar J1144-4308 at redshift $z=0.83$ was recognised as one of the brightest quasars in the night sky (at 14 mag) only in 2022, despite having been imaged on photographic plates since the 19th century\cite{On22}.

In the 1960s finding the first quasars was driven by radio detections of objects that otherwise appeared like regular nearby stars in our Milky Way Galaxy. %Finding quasars later expanded into an industry too rich in methods to mention them all 
Later quasar surveys employed colour selection, template-fitting and Bayesian methods
\cite{Web95,KX,Wis00,Fan01,Wolf03,Stern05,Ri09,Ive14,Reed17}. 
%It was propelled in sample volume by large-area sky scans by the Sloan Digital Sky Survey\cite{York00,Ri09, Lyke20} (SDSS), the all-sky survey by the Widefield Infrared Survey Explorer\cite{WISE} ({\it WISE}), and later projects, up to {\it Gaia}'s all-sky low-resolution spectroscopy\cite{GaiaDSC}.
But the main barrier for obtaining complete samples has always been contamination of quasar candidate selections with stars from our Milky Way, which appear similar if no discriminating information is available and vastly outnumber true quasars in the extreme regime of the brightest quasars.
%Finding quasars started with heuristic approaches as knowledge of the properties of the full quasar population had still to be uncovered prior to becoming useful for informing search criteria. After quasar properties became better understood and easier to model quantitatively, Bayesian probabilistic methods have been used to create more advanced samples\cite{Wolf03,Ri09,Reed17}. 
With a million quasars known by now and the wealth of data from modern all-sky surveys, machine-learning approaches are now most popular\cite{Cal19,Yang23,Quaia}.
%, combining the benefits of large training samples with the available rich characterising data of new query samples. However, %depending on the purpose anticipated and the tuning employed, 
These tend to get the classification right for the majority of objects, while training samples imply that they perform less well for rare extreme cases. In the case of extremely luminous quasars, the obvious bias of a training sample is simply that they do not seem to exist until they are found. 

A low-resolution spectrum of J0529-4351 revealing its quasar nature and redshift has been part of the public all-sky {\it Gaia} DR3 data set published on 13 June 2022. %It is instructive to realise that 
The machine-learning classification of this data set by the {\it Gaia} Discrete Source Classifier (DSC) has assigned J0529-4351 a 99.98\% probability to be a Milky Way star\cite{GaiaDSC}, although a human astronomer eyeballing the {\it Gaia} spectrum would recognise the quasar and redshift at first sight. %In contrast, a large-scale quasar candidate search\cite{Yang23} in the data set of the Dark Energy Survey\cite{DES} (DES) combined with {\it WISE} has assigned to the object a 98\% quasar probability and a photometric redshift estimate of $z_{\rm phot}=3.95^{+0.4}_{-0.2}$.

These days, combining data from {\it Gaia} and {\it WISE} makes an all-sky search for bright quasars straight-forward. An absence of parallax and proper motions removes most of the bright Galactic stars from the search, and the mid-infrared photometry from {\it WISE} is then sufficient to discriminate the disk continuum and dust emission of quasars from the Rayleigh-Jeans tail of stellar photospheres\cite{Stern05}. 
%{\it Gaia} can measure the motions of Milky Way stars easily to a distance of 10~kpc or more. More distant stars may appear motionless like quasars, but when they are bright enough to be considered extreme quasars, they would need to be horizontal-branch stars or red giants. 
%With {\it Gaia} and {\it WISE}, finding quasars brighter than magnitude 17 has become trivial: 
The All-sky Bright Complete Quasar Survey\cite{On23} (AllBRICQS), which reported J0529-4351, selects its candidates with a simple heuristic {\it WISE} colour cut of W1$-$W2$>0.3$ and an absence of $>4\sigma$ evidence for stellar motion. While it aims for {\it completeness} and its recall of known quasars is nearly perfect,
%, its separation of stars and quasars appears good enough to ignore Bayesian or machine-learning approaches, especially given 
its efficiency is still high, given that 97\% of its candidates turn out to be quasars\cite{On23}. 
Given that the AllBRICQS follow-up is complete at $R_p<16$ in both hemispheres, we doubt that a quasar of higher UV-optical luminosity will be found in the future, unless it is hiding behind the Galactic Plane: AllBRICQS did not search at Galactic latitude of $|b|<10\deg$, where source crowding and higher dust extinction makes the discovery of quasars still very challenging\cite{Fu22}.

\section*{Outlook: future observations with bright quasars}

Extremely luminous quasars enable further observations for specific challenging quests: after decades of instrumental development to increase the spatial resolution of telescopic observations, the VLTI/GRAVITY Collaboration\cite{VLTGRAV18} made headlines with their spatio-kinematic mapping of the broad-line region (BLR) in the iconic nearby quasar 3C~273 (redshift $z=0.157$). This observation revealed the orbital rotation of the disk-like BLR in 3C~273 in a spatially resolved pattern and thus provided a direct measurement of the black hole mass from the BLR orbits. The measurement was possible as 3C~273 was bright enough and displayed its Paschen-$\alpha$ emission line in GRAVITY's K-band window. 

While the redshift of J0529-4351 imposed challenges for the work presented here, as nearly all strong UV emission lines are in places heavily affected by atmospheric absorption, it does place its H$\beta$ line at 2.4$\mu$ in the K-band window of the GRAVITY instrument. We predict the size of the H$\beta$-emitting broad-line region in J0529--4351 by extrapolating the radius-luminosity relation\cite{Be13} and find a radius of 2.2~pc. This implies an angular diameter of 0.64~milli-arcseconds, which is an order of magnitude larger than the BLR in 3C~273 and thus the largest-appearing quasar BLR in the Universe. We expect the soon-to-be-upgraded VLTI/GRAVITY+ to obtain a superbly well-resolved picture of the BLR rotation in J0529-4351 and thus a much more reliable measurement of its black hole mass. Crucially, black-hole masses at the high end are estimated from relations that have been extrapolated by orders of magnitude such that the whole scale for objects like this is at risk. Getting a direct mass measurement for a black hole with likely 50$\times$ the mass of the black hole in 3C~273 would be extremely valuable for constraining the relations commonly used to estimate the masses of black holes in the early Universe. The true mass scale of the earliest SMBH would also impact the question of how hard exactly it is to form them.

Future plans include watching the Universe expand with repeat observations of quasars lasting for a decade: the expansion shifts the redshift of individual gas clouds observed as Lyman-$\alpha$ forest absorption lines in quasar spectra\cite{San62,Liske08,Cristiani23} (Sandage test). The signal of such observing campaigns depends on the availability of a sufficient number of very bright quasars and will still require advanced facilities such as the forthcoming ESO Extremely Large Telescope (ELT). J0529--4351 will quite obviously be an important part of this long-term endeavour.

\clearpage

\section*{Methods}

\subsection*{Evidence for point-source geometry}

While {\it Gaia} provides no images as such, we use the source characterisation from its Data Release 3 to provide evidence for a point-source geometry in J0529-4351. First, there is no noticeable flux outside an aperture with 0.175~arcsec radius around the object centroid in the high-resolution imaging\cite{GaiaEDR3}, as determined by comparing the fluxes between $B_P/R_P$ and the $G$-band aperture\cite{On23}. The closest neighbouring object is 3.2~arcsec away; it is not detected by {\it Gaia}, but by the DECam component of the DESI Legacy Imaging Surveys\cite{LSDR9}, with $g=23.581 \pm 0.039$, $r=22.319 \pm 0.015$, $i=20.686 \pm 0.007$, and $z=20.099 \pm 0.007$ in DR10, which is a good match to the colours of an M4 star in the Milky Way. 

Second, astrometric excess noise is absent in J0529-4351, although 319 good observations were available to determine astrometric solutions. All Gaia parameters that may hint at unresolved multiple sources, including \verb|frac_multipeak| (0), \\
\verb|astrometric_excess_noise| (0), \verb|astrometric_sigma5d_max| (0.069), \\
\verb|ruwe| (1.013),  and \verb|ipd_gof_harmonic_amplitude| (0.0167) are in line with the average values (0.132, 0.042, 0.074, 1.015, 0.045) for bright quasars with $R_p<16$ at redshift $z>2$ and far below any selection cuts used to search for Gaia-unresolved binary stars\cite{Fabricius21} and dual or lensed quasars\cite{Chen22}. Figure 1 shows the astrometric excess noise (AEN) of quasars at $R_p<18$ including all dual and lensed quasars\cite{Chen22,Lemon23} with image separations below 0.5 arcsec from the list at \url{https://research.ast.cam.ac.uk/lensedquasars/}, that are unresolved by Gaia; even at the smallest known image separation of 0.18 they show an (AEN)$>1.0$~mas. 

For the lensing calculation we followed a procedure used in the analysis in a recently discovered high-redshift lensed quasar\cite{Yue22}, while adjusting the source redshift to the $z\sim 4$ of J0529-4351. 
%(e.g., \verb|astrometric_excess_noise| $>1.0$).
%, \verb|ipd_gof_harmonic_amplitude| $>0.1$ and \verb|ruwe| $>1.4$).

\subsection*{Spectroscopic observations and data processing}

%The optical spectrum of J0529-4351 was obtained with the Wide Field Spectrograph\cite{Dopita10} (WiFeS) on the ANU 2.3m-telescope at Siding Spring Observatory on 2022, Nov 21. We used the B3000 and R3000 gratings in the blue and red arm, which cover the wavelength range from 3600~\AA \ to 9800~\AA \ at a resolution of $R=3000$. The exposure time was X00~sec and observing conditions ... The data were reduced using the Python-based pipeline PyWiFeS\cite{Childress14}. PyWiFeS calibrates the raw data with bias, arc, wire, internal-flat and sky-flat frames, and performs flux calibration and telluric correction with standard star spectra. Flux densities were calibrated using several standard stars throughout the year, which are usually observed on the same night. We then extract spectra from the calibrated 3D cube using a bespoke algorithm for fitting and subtracting the sky background after masking sources in a white-light stack of the 3D cube. 

The optical/near-IR spectrum of J0529--4351 was obtained with the X-Shooter instrument\cite{2011A&A...536A.105V} on ESO's Very Large Telescope (UT3) on 2023, Jan 14. The airmass at the time of observation was 1.06 and the estimated seeing was about 1.5~arcsec. In the UVB, VIS, and NIR arms, the slit widths and total exposure times were: 1.0~arcsec and 500~s; 0.9~arcsec and 400~s; and 0.9~arcsec and 800~s, respectively. 

The data were reduced using PypeIt, %(\url{https://github.com/pypeit/PypeIt}), 
a Python-based spectroscopic data reduction pipeline \citep{pypeit:joss_pub}. 
%\citep{2019zndo...3506873P,2020arXiv200506505P}.
Because it was not possible to {\it a priori} know the underlying true shape of the \civl\ emission line beneath the atmospheric A-band absorption, we could not reliably fit the telluric model to the J0529--4351 VIS spectrum alone. To obtain an improved correction, we took the telluric model generated from a standard star observed on the same night and modified it by adopting a power-law intrinsic stellar spectrum across a broad span of wavelengths unaffected by strong stellar atmospheric lines. Applying this modified model to J0529--4351 yielded a final spectrum with a smooth \civl\ profile containing only narrow absorption lines and no spurious emission from overcorrected telluric absorption.

The observed spectrum was corrected for slit losses by calibration to external photometric data, in particular, the quasar's VHS DR6 $J$-band magnitude of $14.812\pm0.003$~mag (Vega). With the VHS calibration anchoring the NIR arm, the VIS and UVB data were sequentially matched to the spectroscopic flux levels in the wavelength regions of overlap between the arms. This $J$-band calibration is consistent (within the photometric errors) with that of the $i_{\rm SDSS}$ X-Shooter acquisition image, as calibrated by synthetic photometry for a neighbouring, non-variable star with {\it Gaia} low-resolution spectroscopy, and the flux scale measured in the {\it Gaia} spectrum of J0529--4351 itself.
%SkyMapper Southern Survey (SMSS) DR3 magnitudes in $r$- and $i$-band for the VIS arm; and 2MASS $J$-, $H$-, and $K$-band magnitudes for the NIR arm. The UVB arm was then matched in flux to the overlapping wavelength regime with the VIS arm. This flux calibration agrees to better than 10\% with the flux measured in the {\it Gaia} low-resolution spectrum.

\subsection*{Spectral decomposition and emission-line fitting}

The reduced and telluric-corrected X-Shooter spectrum is then transformed into the rest-frame using a redshift of $z = 3.962$. We model the broad emission-line profile of both the \civl\ line and \mgiil\ line using the code \texttt{PyQSpecFit}\cite{PyQSpecFit_v1}, a python-based spectral modelling package that is designed for rest-frame UV and optical quasar spectra. We model each line individually, selecting independent windows in wavelength to constrain the quasar continuum and the emission-line flux. We mask the narrow absorption features present in our spectrum by applying a boxcar sigma-clipping routine with a width of 50 pixels and a 3$\sigma$ threshold.

For \civl, we constrain the continuum on either side of the feature with a power-law model over line-free wavelength windows of 1445\AA--1455\AA\ and 1973\AA--1983\AA. The emission-line model is composed of a maximum of three broad Gaussian components with a minimum full-width at half-maximum (FWHM) of 1000 km s$^{-1}$. We fit the \civl\ profile between 1480\AA--1528\AA\ and 1537\AA--1565\AA, avoiding the narrow absorption feature between 1528\AA--1537\AA. We also measure the monochromatic luminosity at 1450\AA\ using the power-law model of the continuum. 
We estimate the errors in the spectral fitting by creating 50 realisations of the spectrum after randomly redistributing the flux data according to each point's Gaussian uncertainties.
%We derive the errors by resampling the spectrum 50 times and performing the same spectral modelling routine over the synthetic spectra. We resample the spectrum to create synthetic spectra by randomly selecting a flux value at each pixel assuming that the noise follows a Gaussian probability distribution. 
The standard deviation in each of the derived properties of the \civl\ emission-line is adopted as the statistical error. 

For \mgiil, the continuum model is composed of a power-law and a template of the UV flux contribution from broad \feii\ emission, which is constrained across the wavelength ranges 1973\AA--1983\AA, 2060\AA--2340\AA, 2600\AA--2740\AA, 2840\AA--3100\AA. The emission-line model for \mgiil\ is composed of up to one narrow and three broad Gaussian components, where the narrow and broad distinction is set at a FWHM of 1000 km s$^{-1}$. We fit the \mgiil\ line between 2700\AA--2870\AA\ and the monochromatic luminosity as 3000\AA\ is also measured from the continuum power-law model. We derive the uncertainty in each of the line properties by adopting four different templates of the UV \feii\ emission\cite{Vestergaard_2001,Tsuzuki_2006,Mejia-Restrepo_2016,Bruhweiler_Verner_2008}. As with \civl, we also resample the spectrum 50 times and measure the standard deviation in the \mgiil\ line properties, summing in quadrature with the uncertainty derived from \feii\ templates. More information about the \texttt{PyQSpecFit} modelling of \civl\ and \mgiil\ lines are detailed in a study of the now second-most luminous quasar, SMSS~J2157--3602\cite{Lai23}, and also in a study of line properties in the luminous quasar sample of the European Southern Observatory Large Programme XQ-100\cite{Lai23b}, which is at a comparable redshift range to J0529--4351.

\subsection*{Luminosity calculations}

For calculating monochromatic continuum luminosities, we corrected the spectra for extinction by dust in the Milky Way. We used the estimate of $E(B-V)=0.041\pm 0.0027$ from the Schlegel maps\cite{SFD98} with the correction factor\cite{Sc11} of 0.86 and the extinction law by Fitzpatrick\cite{Fi99}, which yields restframe absorption values of $A_{1450}=0.074$~mag and $A_{3000}=0.025$~mag. We then apply bolometric corrections for mean quasar SEDs\cite{Ri06b} of $k_{1350}=3.81$ and $k_{3000}=5.15$ and apply an anisotropy correction factor of 0.75 assuming mean orientation\cite{Runnoe2012}). The accretion disc continuum fitting described below determines the bolometric luminosity directly from the integrated disc model SEDs and implies a spin- and inclination-dependent anisotropy correction. We then take the average of the luminosity measurements.

%Using mid-infrared data from the Widefield Survey Explorer\cite{WISE} ({\it WISE}), we can further estimate a luminosity of $M=-32.0\pm0.1$ at restframe $4\mu$, which is assumed to be mostly thermal emission from hot nuclear dust. 

%From CAO: There is nothing significant visible in either the RACS-Low or RACS-Mid Epoch1 maps. The closest RACS-Low DR1 source is 2arcmin away. The median noise is 250 microJy/beam (McConnell et al. 2020, PASA, 37, e048).

\subsection*{Spectral decomposition and continuum fitting}

We use a publicly available code\cite{BADFit_v1} to model the shape of the accretion disc continuum to spectral energy distribution models predicted for slim discs with \texttt{slimbh}\cite{Sad11}, which is a grid of synthetic spectra from ray-traced numerical solutions of slim accretion disc equations. The free parameters are mass and spin of the black hole as well as accretion rate and disc inclination. A Markov chain Monte Carlo (MCMC) method is used to map out likelihood contours across the 4D parameter space. 
We collect the Milky Way extinction-corrected spectrum and extend the IR coverage with {\it{WISE}} photometry, specifically the CatWISE2020\cite{Marocco_2021_CatWISE2020} \textit{W1} and \textit{W2} photometric passbands. We then create synthetic data points to represent the accretion disc flux within selected line-free windows in the observed spectrum. For the \textit{WISE} wavelengths, we adopt the spectrum of the Selsing high-luminosity quasar template\cite{Selsing_2016}, scaled independently to the flux of \textit{W1} and \textit{W2} bandpasses, and create synthetic line-free data points in the same fashion. Furthermore, we use the X-Shooter UVB arm to estimate the continuum flux through the Ly$\alpha$ forest and create one more synthetic data point to constrain the continuum. We then use MCMC to infer the Bayesian posterior probability distributions of the intrinsic black hole properties using the set of synthetic data points to represent the flux of the accretion disc continuum. 

Further detailed information on the continuum fitting method is presented in an in-depth study of the quasar SMSS J2157--3602\cite{Lai23}. However, unlike this previous study, the static and smooth thermal accretion disc models are unable to fully reproduce the hardening of the continuum in the VIS arm or the flux passing through the Ly$\alpha$ forest. 
%The recent brightening as seen in the light curve may be responsible for preferentially increasing the shorter wavelength flux as the hotter inner parts of the disc respond more rapidly and more strongly to the enhanced nuclear heating. Another possibility is that 
It has been shown previously that quasars with high \civl\ blueshift, which are indicative of strong outflows and winds, tend towards a bluer UV continuum to the blue side of $\lambda_{\rm rest}\approx 200$~nm and a slightly redder continuum on the red side\cite{Temple_2021}, thus producing a spectral break that is missing in the thermal disc models. Therefore, our \texttt{slimbh} synthetic spectra utilises the disc atmosphere model, \texttt{BHSPEC} \cite{Davis_2005, DH_2006_BHSPEC}, to help reproduce the emerging Compton-hardened radiation.

We also attempted to fit the data with alternative models. E.g., we adopt \texttt{kerrbb} thin disc models\cite{Li05} despite the fact that the slim disc model reproduces the thin disc SED at low Eddington ratios, and find indeed similar results for the mass of the black hole and the luminosity of the disc, although the colour of the UV continuum is not properly reproduced in the absence of Comptonisation considerations. We also examine the option of J0529--4351 being lensed by demagnifying the spectrum and find that it does not improve the quality of fit. 
%Therefore, we take the original result from the slim disc but due to the departure of the observed SED from the static thermal model, we place no additional weight on it relative to the two single-epoch virial mass estimates for the final, averaged black hole mass measurement.

\subsection*{Host galaxy dust extinction}

We assume no dust extinction in the quasar host galaxy given that typical host reddening levels are found to be consistent with $E(B-V)\approx 0.0$ among luminous quasars\cite{Kr15}, while less than 1\% of quasars seem to have $E(B-V)>0.1$. Also, an object appearing as the brightest object in the Universe has a low probability of being extinguished by notable levels of dust. If dust were present, it would make the continuum redder and the measured luminosity lower. This will lead to the black hole mass being overestimated by the continuum-fitting method and underestimated by the virial single-epoch method. Such a discrepancy is not observed here.

%\subsection*{Black-hole mass calibrations}

%CIV calibrations\cite{Vestergaard_2006, Coatman_2017}
%MgII calibrations\cite{Vestergaard_2009, Shen_2011, Le_2020}

\subsection*{Light-curve construction}

The light curve with photometry from the NASA (National Aeronautics and Space Administration) Asteroid Terrestrial-impact Last Alert System\cite{To18a} (ATLAS) 0.5m telescope was obtained from the ATLAS website. For any object in the ATLAS observing area, up to four observations per night are available, depending on weather. The orange passband is observed in all clear nights, while the cyan passband is only used during the half period around New Moon. For slowly varying objects, noise can be suppressed by combining observations from longer periods; we determine median measurements to reduce the influence of outlier measurements, using per-week intervals in the orange passband and per-moon period intervals in the cyan passband. Error bars on the medians express the level of variability within the interval by showing inter-quartile ranges of the values.

\section*{Data availability}
Data from Data Release 3 (DR3) of the European Space Agency's {\it Gaia} mission are publicly available (\href{https://gea.esac.esa.int/archive/}{https://gea.esac.esa.int/archive/}). NASA ATLAS data are available from \href{https://fallingstar-data.com/forcedphot/}{https://fallingstar-data.com/forcedphot/}. The SkyMapper Southern Survey data are available from \href{https://skymapper.anu.edu.au/}{https://skymapper.anu.edu.au/} (doi:10.25914/5f14eded2d116). The raw spectrum and calibration files from the ESO/VLT are available in the ESO archive at \href{http://archive.eso.org/}{http://archive.eso.org/}. A reduced spectrum is available from the authors on reasonable request.

\section*{Code availability}
The spectral fitting code and the quasar continuum fitting code were written by SL in python and are publicly available on github at \href{https://github.com/samlaihei/PyQSpecFit}{https://github.com/samlaihei/PyQSpecFit}\cite{PyQSpecFit_v1} and \href{https://github.com/samlaihei/BADFit}{https://github.com/samlaihei/BADFit}\cite{BADFit_v1}.

\section*{Acknowledgements}
This work was supported by the Australian Research Council (ARC) through Discovery Project DP190100252 (CW, FB, CAO, SL). SL is grateful to the Research School of Astronomy \& Astrophysics at Australian National University for funding his Ph.D. studentship. We thank Giovanni Ferrami from the University of Melbourne for discussing solutions for strong gravitational lensing. %We acknowledge the traditional owners of the land on which the telescopes of Siding Spring Observatory stand, the Kamilaroi people, and pay our respects to their elders, past and present.

Data for this project were obtained at the European Southern Observatory through DDT proposal 2110.B-5032.

This work has made use of data from the European Space Agency mission {\it Gaia} (\href{https://www.cosmos.esa.int/gaia}{https://www.cosmos.esa.int/gaia}), processed by the {\it Gaia} Data Processing and Analysis Consortium (DPAC,\href{ https://www.cosmos.esa.int/web/gaia/dpac/consortium}{ https://www.cosmos.esa.int/web/gaia/dpac/consortium}). Funding for the DPAC has been provided by national institutions, in particular the institutions participating in the {\it Gaia} Multilateral Agreement.

This publication makes use of data products from the {\it Wide-field Infrared Survey Explorer}, which is a joint project of the University of California, Los Angeles, and the Jet Propulsion Laboratory/California Institute of Technology, and NEOWISE, which is a project of the Jet Propulsion Laboratory/California Institute of Technology. {\it WISE} and NEOWISE are funded by the National Aeronautics and Space Administration.

SuperCOSMOS Sky Survey material is based on photographic data originating from the UK, Palomar and ESO Schmidt telescopes and is provided by the Wide-Field Astronomy Unit, Institute for Astronomy, University of Edinburgh.

This work made use of Astropy (\href{http://www.astropy.org}{http://www.astropy.org}), a community-developed core Python package and an ecosystem of tools and resources for astronomy \citep{%astropy:2013, astropy:2018, 
astropy:2022}.

This research has made use of the SVO Filter Profile Service (\href{http://svo2.cab.inta-csic.es/theory/fps/}{http://svo2.cab.inta-csic.es/theory/fps/}) supported from the Spanish MINECO through grant AYA2017-84089.

The national facility capability for SkyMapper has been funded through ARC LIEF grant LE130100104 from the Australian Research Council, awarded to the University of Sydney, the Australian National University, Swinburne University of Technology, the University of Queensland, the University of Western Australia, the University of Melbourne, Curtin University of Technology, Monash University and the Australian Astronomical Observatory. SkyMapper is owned and operated by The Australian National University's Research School of Astronomy and Astrophysics. The survey data were processed and provided by the SkyMapper Team at ANU. The SkyMapper node of the All-Sky Virtual Observatory (ASVO) is hosted at the National Computational Infrastructure (NCI). Development and support of the SkyMapper node of the ASVO has been funded in part by Astronomy Australia Limited (AAL) and the Australian Government through the Commonwealth's Education Investment Fund (EIF) and National Collaborative Research Infrastructure Strategy (NCRIS), particularly the National eResearch Collaboration Tools and Resources (NeCTAR) and the Australian National Data Service Projects (ANDS).

This work uses data from the University of Hawaii's ATLAS project, funded through NASA grants NN12AR55G, 80NSSC18K0284 and 80NSSC18K1575, with contributions from the Queen's University Belfast, STScI, the South African Astronomical Observatory and the Millennium Institute of Astrophysics, Chile.

\section*{Author Contributions Statement}
All authors contributed to data collection. SL led the data analysis with contributions from CAO and CW. CW selected the quasar candidates and led the drafting and editing of the article.

\section*{Competing Interests Statement}
The authors declare no competing interests.

\clearpage

\begin{table}
	\begin{center}
	\caption{Luminosity, black hole mass and accretion rate for the quasar J052915.80--435152.0: given are best-fit estimates and 68\% confidence intervals. Systematic uncertainties for the mass estimates are proposed to be as high as 0.4~dex. }
    \vspace{3mm}
	\label{tab:masstab}
	\begin{tabular}{ll}
		\hline
        %Method          
		\hline
        {Luminosity} &   log $L/(\mathrm{erg~s}^{-1})$ \\
        \hline
        $\lambda L_{135}$   &  $47.93$ \\
        $\lambda L_{300}$   &  $47.76$ \\
		\hline
        $L_{\rm bol}$ (from 135 nm)     &  $48.39$ \\
        $L_{\rm bol}$ (from 300 nm)     &  $48.35$ \\
        $L_{\rm bol}$ (slim disc model) &  $48.16$  \\
        $L_{\rm bol}$ (best estimate)   &  $48.27\pm 0.06$ \\  % 0.02 dex down
		\hline
		%\hline
        {Black hole mass}    & log $M/M_\odot$ \\
        \hline
		%\civl\ line 	(Coatman+\cite{Coatman_2017})  & $10.03\pm 0.06$	\\
		%\civl\ line 	(VP06\cite{VP06})       & $10.45\pm 0.02$	\\
        \civl\ line 	(average)               & $10.24\pm 0.15$  \\
        %\hline
		%\mgiil\ line 	(VP09\cite{Vestergaard_2009})  & $10.03\pm 0.09$	\\
		%\mgiil\ line 	(LWX\cite{Le_2020})     & $10.21\pm 0.09$	\\
		%\mgiil\ line 	(Shen+\cite{Shen_2011}) & $10.36\pm 0.09$	\\
        \mgiil\ line 	(average)               & $10.20\pm 0.08$  \\
		%\hline
%        continuum fit & thin disc\cite{Li05}   & $ 9.81^{+0.18}_{-0.17}$ \\
        Continuum fit  (slim disc\cite{Sad11})  & $10.28^{+0.17}_{-0.10}$ \\
        All methods (best estimate)             & $10.24\pm 0.02$  \\
		\hline
		%\hline
        {Mass accretion rate}    & log $\dot{M}/(M_\odot\mathrm{~yr}^{-1})$ \\
        \hline
		%\civl\ line 	(Coatman+\cite{Coatman_2017})  & $10.03\pm 0.06$	\\
		%\civl\ line 	(VP06\cite{VP06})       & $10.45\pm 0.02$	\\
        for plausible inclination range         & $2.57\pm 0.06$  \\
		\hline
	\end{tabular}
	\end{center}
\end{table}

\clearpage

\clearpage

%Easier citing method. Add new reference to Variability_of_QSOs.bib. After the paper is done, download the .bbl file from overleaf, then paste everything there under \begin{thebibliography}{}. Comment out \bibliographystyle and \bibliography in the end
%\bibliographystyle{plainnat}
%\bibliographystyle{pasa-mnras}
%\bibliographystyle{unsrtnat}
%\bibliographystyle{rusnat}
%\bibliographystyle{naturemag}

%\bibliographystyle{naturemag-cus}
%\bibliography{CW_QSOs}

\end{document}